\documentclass[review]{elsarticle}

\usepackage{lineno,hyperref}
\modulolinenumbers[1]

\usepackage[utf8]{inputenc}

\title{Shear-Induced Activation and Transport of Platelets in Artificial Heart Valve Flows}

\usepackage{geometry}
\usepackage{amsmath}

\usepackage{adjustbox}
\usepackage{xcolor}

\journal{Arxiv}

\begin{document}
\author[1]{Syed Samar Abbas}
\author[1]{Iman Borazjani\corref{cor1}}
\ead{iman@tamu.edu}
\cortext[cor1]{Corresponding author}
\fntext[fn1]{This research was partially funded by Novostia, SA. The computational resources were provided by the High-Performance Computing Research (HPRC) facility at Texas A\&M University.}

\affiliation[1]{J. Mike Walker '66 Department of Mechanical Engineering, Texas A\&M University, College Station, USA}

\begin{abstract}
 
Albeit the hemodynamics of artificial heart valves has been investigated for several decades, the local shear-induced activation potential and subsequent transport phenomena of activated platelets in different valve designs, which mediate thrombosis, remains poorly understood. Here, platelet activation due to local shear stresses and the associated transport phenomena are investigated in two designs of mechanical heart valves (MHVs), namely a trileaflet MHV (TMHV) and a bileaflet MHV (BMHV) and compared against a surgical bioprosthetic heart valve (BHV) as a control. It is observed that the local activation and transport of platelets in any aortic region reach a cyclic state, with MHVs showing higher levels of both activation and transport than BHV. When integrated over the volume of the aortic sinuses and central lumen, the local activation is, respectively, 5.90 and 2.26 times higher in BMHV whereas 2.97 and 1.39 times higher in TMHV than in BHV. The washout of activated platelets from the sinuses and central lumen is, respectively, 10.40 and 2.39 times higher in BMHV while 4.90 and 1.40 times higher in TMHV compared to BHV. The low washout of sinuses in BHV is also demonstrated by higher residence time in sinuses compared to MHVs. These findings indicate that the risk of clinical thrombosis in MHVs is likely due to higher levels of local shear-induced activation than BHV despite the lower residence time (i.e. a better washout). Conversely, the subclinical thrombosis in BHVs is probably due to prolonged platelet residence time relative to MHVs.
    
\end{abstract}


\maketitle

\section{Introduction}

Bileaflet mechanical heart valves (BMHVs) are known for their non-physiological hemodynamics, primarily characterized by supraphysiological shear stresses, vortex shedding and flow separation \citep{sotiropoulos2009review,sotiropoulos2016fluid,abbas2020numerical,abbas2022state}. Several \emph{In Vitro} studies of BMHVs have evidenced a strong correlation between their hemodynamics and platelet activation potential \citep{yin2004flow,alemu2010design,bark2017hemodynamic}. Once activated, platelets can: (a) release von Willebrand factor (vWF) and adenosine diphosphate (ADP) which activates more platelets \citep{casa2017thrombus} and (b) promote the formation of thrombin on their procoagulant surfaces \citep{fogelson2015fluid}. Depending on the local shear rate, the vWF or thrombin (via fibrinogen) can stabilize the platelet plug \citep{rubenstein2018platelet}. Shear-induced platelet activation may act as a precursor to clotting events in BMHVs, requiring lifelong anticoagulation therapy with a vitamin K antagonist (like warfarin) for its recipients \citep{jaffer2019blood}.

To quantify shear-induced platelet activation in BMHVs, numerical Fluid-Structure Interaction (FSI) simulations have been coupled with various empirical models, e.g.  %These models mainly include 
the Linear model \citep{hellums19941993}, the Damage Accumulation models  \citep{yeleswarapu1995mathematical,nobili2008platelet}, and the Soares model \citep{soares2013novel}. Bluestein et al. \citep{bluestein2000vortex,bluestein2002free} reported that platelets could be activated by a combination of high shear exposure around BMHV leaflets and entrapment in the ensuing vortex street. They hypothesized that BMHV's wake might provide favorable conditions for emboli formation, which can convect to the distal arteries with the transport of shed vortices. Similar observations were made by Ahmed et al. \citep{ahmed2022ramifications}, who demonstrated that high vorticity regions in BMHV flows tend to contain more platelets with elevated activation levels compared to low vorticity regions. Alemu and Bluestein \cite{alemu2007flow} noted that implantation misorientation of a BMHV promotes flow separation and formation of counter-rotating spiral vortices during systole, which draw platelets from the core flow to the high shear regions near solid boundaries. Morbiducci et al. \cite{morbiducci2009blood} reported that streamwise vorticity is more strongly correlated with platelet activation in a BMHV relative to spanwise vorticity. Yun et al. \cite{min2014blood} found that vortex shedding anew the BMHV leaflets alone does not significantly activate platelets. Instead, they argued that platelet activation is more likely in the mixing region where sinus recirculation interacts with the vortex street. They also noted that leakage jets during diastole pose a lower risk than persistent recirculation of platelets near the valve. However, Dumont et al. \cite{dumont2007comparison} identified regurgitation during diastole as the most critical cardiac phase for platelet activation in BMHVs, owing to leakage jets squeezing into the left ventricle through small openings in the valve. Abbas et al. \cite{abbas2020num} also reported that the increased severity of BMHV's implantation misorientation leads to intensified platelet activation due to enhanced intensity of intermittent regurgitation.

A major limitation of the reviewed studies is their use  of Lagrangian frameworks, tracking only a small number of `platelet-like' particles far below the physiological range (150,000–450,000 per $mm^3$ of blood). Conjugated with uncertainty around platelet seeding locations, this approach involves concerns about the sensitivity of the reported activation metrics. To alleviate this problem, Hedayat et al. \cite{hedayat2017platelet} developed an Eulerian framework and reported that BMHV generates several-fold higher platelet activation than BHV during systole. Extending this framework for a left ventricle-aorta configuration, Asadi et al. \cite{asadi2022effects} found that variations in implantation rotation of BMHVs have a minimal impact on shear stress and platelet activation. This finding matched the previous reports that a BMHV's rotational misorientation does not alter the shear stress distribution in anatomic aorta but may change the leaflet kinematics \citep{borazjani2010effect,borazjani2010high}.

In efforts to develop a MHV with reduced thrombogenecity, several designs of trileaflet MHVs (TMHVs) have been introduced, offering improved hemodynamics over BMHVs due to their design \citep{vennemann2018leaflet,li2020comparative,abbas2025closure}. Nitti et al. \cite{nitti2022numerical} reported that a TMHV generates shear stresses below the general platelet activation threshold (i.e $<10$ Pa), unlike BMHV (with or without vortex generators). Schubert et al. \cite{schubert2019novel} observed only minor ‘clotting deposits’ on a TMHV compared to a BMHV under similar \emph{In Vitro} conditions, although their use of saline water may not reflect physiological coagulation behavior. To our knowledge, there are no published numerical studies quantifying platelet activation in a TMHV and comparing it with another MHV design or a BHV under similar conditions. It is important to mention here that BHV recipients are also susceptible to the risk of clotting as cases of subclinical thrombosis have been reported in such valves following both surgical \cite{rashid2023computed} and transcatheter implantation \cite{puri2017bioprosthetic}. Moreover, the relative significance of local shear-induced activation versus the transport mechanisms, i.e. the accumulation and/or washout of activated platelets from/to a region of different concentration in these valves have yet to be characterized.

In this research, we quantify the platelet activation levels in a TMHV and a BMHV, and compare them against a typical surgical BHV as a control, under similar numerical conditions. We also evaluate the relative significance of the local shear-induced activation of platelets versus their transport in multiple regions of interest by comparing the contribution of activation source (derived from shear stresses) and accumulation/washout towards the content concentration of activated platelets. Through this analysis, we aim to provide insights that can inform the design of artificial heart valves with reduced platelet activation potential and consequently lower risk of clinical or subclinical thrombosis, thereby improving their long-term clinical performance. 

\section{Methodology}

This study compares shear-induced platelet activation levels in the TRIFLO Valve (Novostia, Switzerland) and the On-X Valve (Artivion, USA), modeled as the TMHV and BMHV, respectively, and a typical surgical BHV, modeled as an assembly of three elastic leaflets that replicate the stress-strain behavior of a natural/tissue valve \citep{borazjani2013fluid,asadi2023contact}. The BHV has been chosen as a control due to its excellent hemodynamic performance and less thrombogenicity, similar to previous work \citep{hedayat2017platelet}. Platelet activation is computed based on our previously simulated flow field for these valves under similar conditions \citep{abbas2025closure} using the curvilinear immersed boundary (CurvIB) method \citep{ge2007numerical,borazjani2008curvilinear}. 
Below, we describe the details of the method that quantifies shear-induced platelet activation and simulates their transport in artificial heart valve flows. The governing equation is a 3D convection-diffusion-production equation which, in its non-dimensional form, reads as follows: 

\begin{equation}
\displaystyle\frac{\partial P}{\partial t} + \nabla \cdot(\vec{u}P)= \frac{1}{Pe}\nabla^2 P + S(\tau)
  \label{governing-eq}
\end{equation}

\noindent where $P$ is the platelet activation level, $\vec{u}$ is the velocity of the flow, and $S(\tau)$ is the production/activation source term which depends on the scalar viscous shear stresses, $\tau$. The Peclet number (\textit{Pe}) in Equation \ref{governing-eq} is defined as: 

\begin{equation}
Pe = \frac{LU}{D_p}
  \label{peclet-num}
\end{equation}

\noindent where, $D_p$ is the diffusivity of platelets, set as $2.5 \times 10^{-5} ~mm^2/s$ \citep{montgomery2023clotfoam}, $L$ is the characteristic length and $U$ is the bulk velocity, adopted from our previous study \citep{abbas2025closure}, and set to, respectively, $25.82~mm$ (the inlet diameter of the aorta) and $654.8~mm/s$.

The production term $S(\tau)$, hereafter referred as the local shear-induced activation, is incorporated from: (a) the Linear model and (b) the Soares model, to test the independence of the drawn conclusions from the activation model used. Both models treat $\tau$ as the mechanical stimulus on resting platelets, not the Reynolds stresses, as characterized by Ge et al. \cite{ge2008characterization}. The scalar shear $\tau$ is calculated as an invariant of the stress tensor as follows \cite{apel2001assessment}:

\begin{equation}
\tau(t)= \left[ \frac{1}{6} \sum_{i,j=1}^3(\sigma_{ii}-\sigma_{jj})(\sigma_{ii}-\sigma_{jj})+\sigma_{ij}\sigma_{ij} \right]^{1/2}
\label{shear-scalar}
\end{equation}

\noindent{where, $\sigma$ is the viscous shear stress tensor, calculated from the strain rate tensor as follows:}

\begin{equation}
\sigma_{ij}= \mu ~(\frac{\partial u_{j}}{\partial x_{i}}+\frac{\partial u_{i}}{\partial x_{j}}),~~~~~(i,j=1~to~3)
\label{shear-viscous}
\end{equation}

The advection term in Equation \ref{governing-eq} is discretized by using the Monotonic Upstream-Centered Scheme for Conservation Laws (MUSCL), coupled with the superbee flux limiter to achieve second-order accuracy while avoiding spurious oscillations \citep{hedayat2017platelet,hedayat2019comparison}. The diffusion term is discretized by using a second-order accurate central differencing scheme. The solution is marched in time by using a total variable diminishing, two-step Runge Kutta method, which has been shown to be second-order accurate \citep{gottlieb1998total}. A Neumann boundary condition ($\frac{\partial P}{\partial n} = 0$) is applied at the walls. For nodes that are just adjacent to immersed bodies, the boundary conditions are reconstructed by enforcing $\frac{\partial P}{\partial n} = 0$ along the normal to boundary, as quintessentially done in the CurvIB method \citep{borazjani2008curvilinear}. 

The Reynold's transport theorem for the conservation of a scalar is used to determine the shear-induced total platelet activation (TPA) in the computational domain as follows:

\begin{equation}
\frac{d}{dt}~\textrm{(TPA)}_{sys}= \frac{\partial}{\partial t} \int_{CV}~P~dV~+~\int_{CS}~\left[P~\vec{u}~.~\vec{n}-\frac{1}{Pe}\nabla P \right]dA=\int_{CV}  S(\tau) dV
  \label{tpa-eq}
\end{equation}

where `CV' refers to the control volume, `sys' to the system, and `CS' to the control surface. $\vec{n}$ is the unit outward normal to the control surface. By integrating over time, the TPA from $t=0$ to $t=T$ can be calculated as:

\begin{equation}
\textrm {TPA}~\Big|_{0}^{T}= \left(\int_{V}~P~dV \right)\Big|_{0}^{T}~+~\int_{0}^{T} \left(~\int_{A}~P~\vec{u}~.~\vec{n}~dA\right)~dt
  \label{eq4}
\end{equation}

\noindent{where, $T$ is the non-dimensional duration of cardiac cycle, such that $T=0.85714/(L/U)=21.7372$. Here, 0.85714s is the duration of the cardiac cycle, corresponding to 70 beats per minute}.

To determine whether the platelet activation level in a spatial region/volume of interest ($V$) is dominated by local shear-induced activation or due to net accumulation characterized by the transport (i.e. importation) of activated platelets from a region of higher activation, we track the volume integral of $P$ (termed $P_V$) from Equation \ref{governing-eq} at time $T$ as follows: 

\begin{equation}
P_V = P_{T_{V}} + P_{P_V} + P_o
  \label{tracking-eq}
\end{equation}

\noindent where $P_o$ is the initial platelet activation level at time $t=0$, being zero for the Linear model whereas 0.01 for the Soares model (i.e. an initial 1\% background level of activation, assumed only for the $1^{st}$ cycle).  $P_{T_V}$ and $P_{P_V}$ are the volume integrals of the time-integrated transport (divergence of fluxes) and local shear-induced activation terms at time $T$, respectively, as follows: 

\begin{equation}
P_{T_V}= -\int_V \left[\int_{t=0}^T\nabla \cdot\left(uP - \frac{1}{Pe}\nabla P\right)dt\right]dV
\label{eq:PTv}
\end{equation}

\begin{equation}
P_{P_V}= \int_V \left[\int_{t=0}^T S(\tau)dt\right]dV
\label{eq:PPv}
\end{equation}

The residence time ($T_{R}$) is calculated as an indicator of flow stasis/stagnation by solving an advection equation for a passive scalar with a constant source term in an Eulerian framework, following past numerical studies \cite{sadrabadi2021fluid,oks2024effect}:

\begin{equation}
\displaystyle\frac{\partial T_{R}}{\partial t} + \nabla \cdot(\vec{u}T_{R})= 1
  \label{residencetime-eq}
\end{equation}

All equations are first transformed into generalized curvilinear coordinates (see \ref{app:num-method}) and then solved over distributed memory nodes under an efficiently parallelized framework by employing MPI and the Portable, Extensible Toolkit for Scientific Computation (PETSc) \citep{balay2019petsc}.

\section{Results}

To ensure that the results reach a cyclic state, we simulated 15 cycles and report the platelet activation results from the $15^{th}$ cardiac cycle in this section. Note that the activation source (derived from $\tau$) is identical for each cardiac cycle since $\tau$ is calculated from the hemodynamically variations-independent cycle established in our previous FSI study \citep{abbas2025closure}. All contours are plotted on the x-midplane and a cross-sectional plane located at the sinotubular junction (STJ). 

%%%%%%%%%%%%%%%%%%%%%%%%%%%%%%%
\begin{figure}[hbtp]
	\begin{minipage}{\textwidth}
		\begin{center}
			\includegraphics[width=1\textwidth]{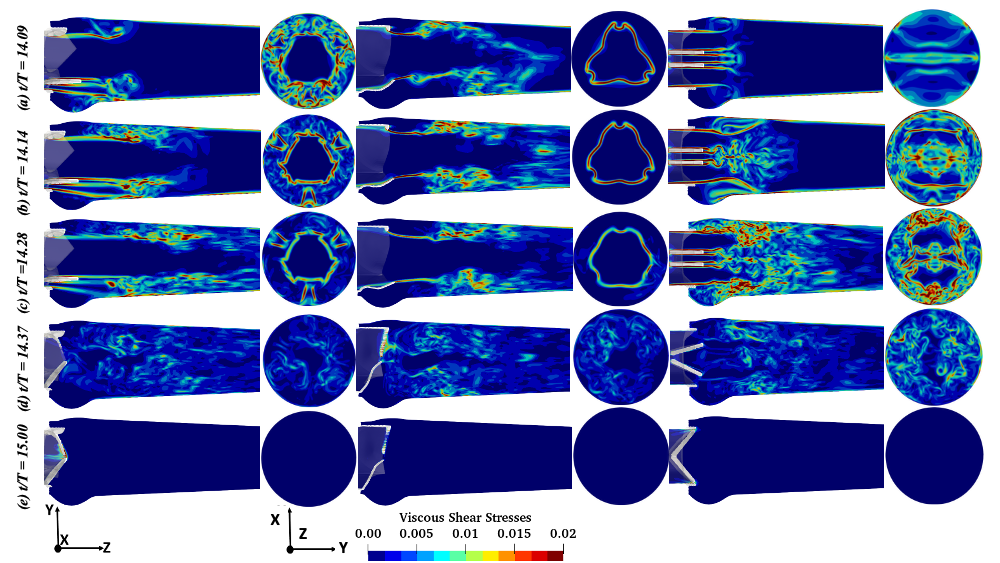}
			\caption{Viscous shear stresses in the TMHV (left), BHV (middle) and BMHV (right) at (a) early systole (b) peak systole (c) late systole (d) early regurgitation and (e) end of diastole. See Supplementary Movie 1.}
			\label{shear}
		\end{center}
	\end{minipage}
\end{figure}
%%%%%%%%%%%%%%%%%%%%%%%%%%%%%%%

We begin by examining the scalar viscous shear stresses in the three valves (Figure \ref{shear}) since the platelet activation patterns are generally considered a strong reflection of shear stress distributions. During systolic acceleration (Figure \ref{shear}a), elevated shear stresses are observed in the shear layers concealing the orificial jets compared to other regions of the domain. By peak systole (Figure \ref{shear}b), the MHVs develop strong shear layers around their leaflets, with shear stresses exceeding 10 Pa, in contrast to BHV. These shear layers extend into the ensuing wake of all valves, but become immediately unstable only for the BMHV by peak systole. Consequently, strong shed vortices are observed anew the BMHV's leaflets, forming a \emph{von Karman}-like vortex street with high shear stresses (Figure \ref{shear}b), persisting through late systole (Figure \ref{shear}c). In contrast, the immediate wakes of the TMHV and BHV remain devoid of vortices and shear due to dominant central jets (Figure \ref{shear}b-c). All valves show localized high shear stresses near the aortic wall beyond the sinus ridge during systole (Figure \ref{shear}b–c), primarily driven by flow reversal in that region \cite{abbas2025closure}, with BMHV demonstrating highest shear stresses at the STJ. Furthermore, BMHV also shows the highest shear stress within the sinuses among all valves (Figure \ref{shear}b–c), consistent with its strongest recirculating flow reported in our previous study \citep{abbas2025closure}.  The shear stresses continue to weaken during early regurgitation (Figure \ref{shear}d), and by the end of diastole, only the leakage jets of the three valves are encapsulated by layers of high shear stresses (Figure \ref{shear}e). As will be shown subsequently, shear stress distributions alone do not directly dictate the spatial patterns of platelet activation.

%%%%%%%%%%%%%%%%%%%%%%%%%%%%%%%

\begin{figure}[hbtp]
	\begin{minipage}{\textwidth}
		\begin{center}
			\includegraphics[width=1\textwidth]{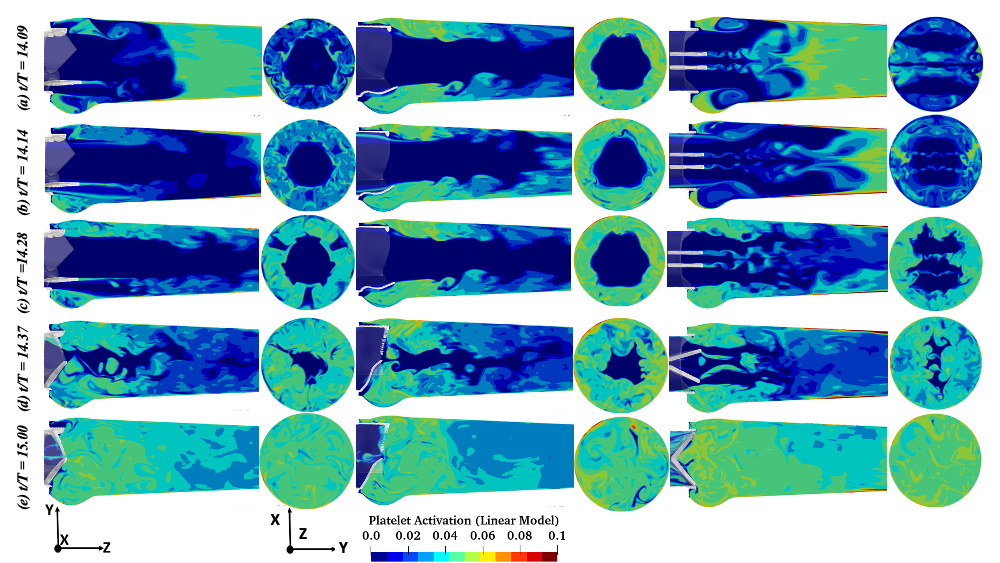}
			\caption{Platelet activation levels (Linear Model) in TMHV (left), BHV (middle) and BMHV (right) at (a) early systole (b) peak systole (c) late systole (d) early regurgitation and (e) end of diastole. See Supplementary Movie 2.}
			\label{linear-cont}
		\end{center}
	\end{minipage}
\end{figure}

\begin{figure}[hbtp]
	\begin{minipage}{\textwidth}
		\begin{center}
			\includegraphics[width=1\textwidth]{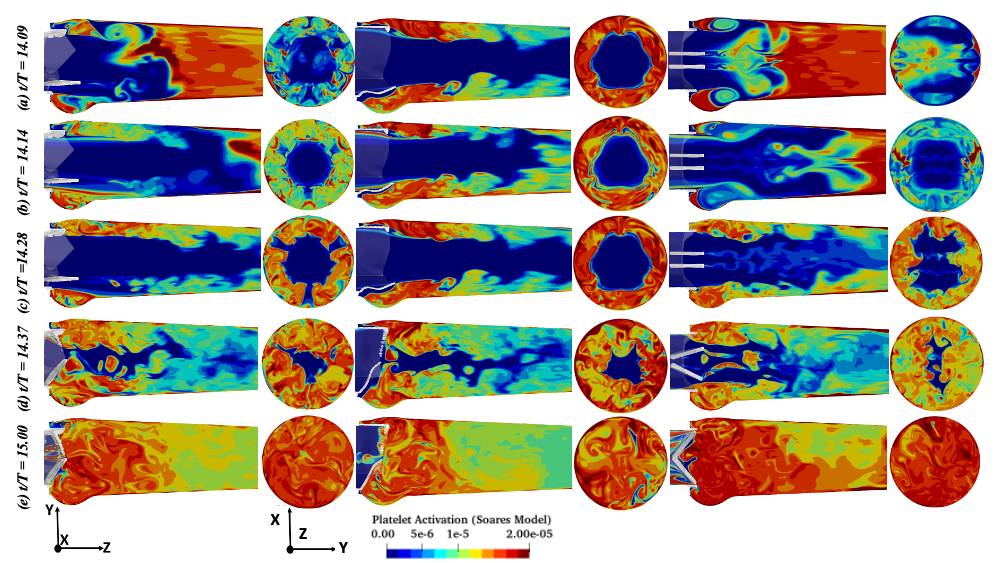}
			\caption{Platelet activation levels (Soares Model) in TMHV (left), BHV (middle) and BMHV (right) at (a) early systole (b) peak systole (c) late systole (d) early regurgitation and (e) end of diastole.
            See Supplementary Movie 3.}
			\label{soares-cont}
		\end{center}
	\end{minipage}
\end{figure}
%%%%%%%%%%%%%%%%%%%%%%%%%%%%%%%%%

Figures \ref{linear-cont} and \ref{soares-cont} show the contours of platelet activation levels obtained from the Linear and Soares models, respectively. The forward orificial jets of all valves transport fresh, non-activated platelets from the inlet and wash away the activated platelets to the downstream regions throughout the systole (Figures \ref{linear-cont}a-c and \ref{soares-cont}a-c). Specifically, both models show that the region covered by the central jet of TMHV and BHV, which was previously shown to be devoid of vortices \citep{abbas2025closure} and shear stresses (Figure \ref{shear}a-c), has low, near-zero platelet activation levels until late systole (Figures \ref{linear-cont}a-c and \ref{soares-cont}a-c). In contrast, the ensuing wake of BMHV leaflets, which was shown to be characterized by strong vortex shedding \citep{abbas2025closure} and high shear stresses (Figure \ref{shear}a-c), generates elevated levels of platelet activation compared to TMHV and BHV. Notably, the MHVs have higher platelet activation levels near the exit of the domain compared to BHV (Figures \ref{linear-cont}a-b and \ref{soares-cont}a-b), demonstrating a higher effectiveness of the central jet washout in the latter. Furthermore, for all valve types, the aortic sinuses and the near-arterial wall region beyond the sinus ridge (i.e. the region between the orificial jets of each valve and the aortic wall) consistently exhibit pronounced platelet activation levels (Figures \ref{linear-cont}a-b and \ref{soares-cont}a-b). In these regions, the lateral jets in the MHVs during the acceleration phase wash out the platelets with high activation levels from the previous cycle, in contrast to BHV, which lacks lateral jets. Consequently, BHV exhibits the highest levels of platelet activation in the sinuses and regions near the arterial walls, followed by the TMHV, and lowest in BMHV (Figures \ref{linear-cont}a-c and \ref{soares-cont}a-c). This trend persists until the complete closure of BHV and TMHV (during early regurgitation), as shown in Figures \ref{linear-cont}d and \ref{soares-cont}d. At the end of diastole (Figures \ref{linear-cont}e and \ref{soares-cont}e), however, the BMHV domain displays the strongest pockets of platelet activation, surpassing both the TMHV and BHV in terms of strength and the area covered. Given the BHV's well-known hemodynamic superiority, the above findings prompt an important question: Does the BHV cause a higher local shear-induced activation relative to MHVs, specifically during systole, or the observed activation patterns are more strongly influenced by transportation dynamics?

%%%%%%%%%%%%%%%%%%%%%%%%%%%%

\begin{figure}[hbtp]
	\begin{minipage}{\textwidth}
		\begin{center}
			\includegraphics[width=1\textwidth]{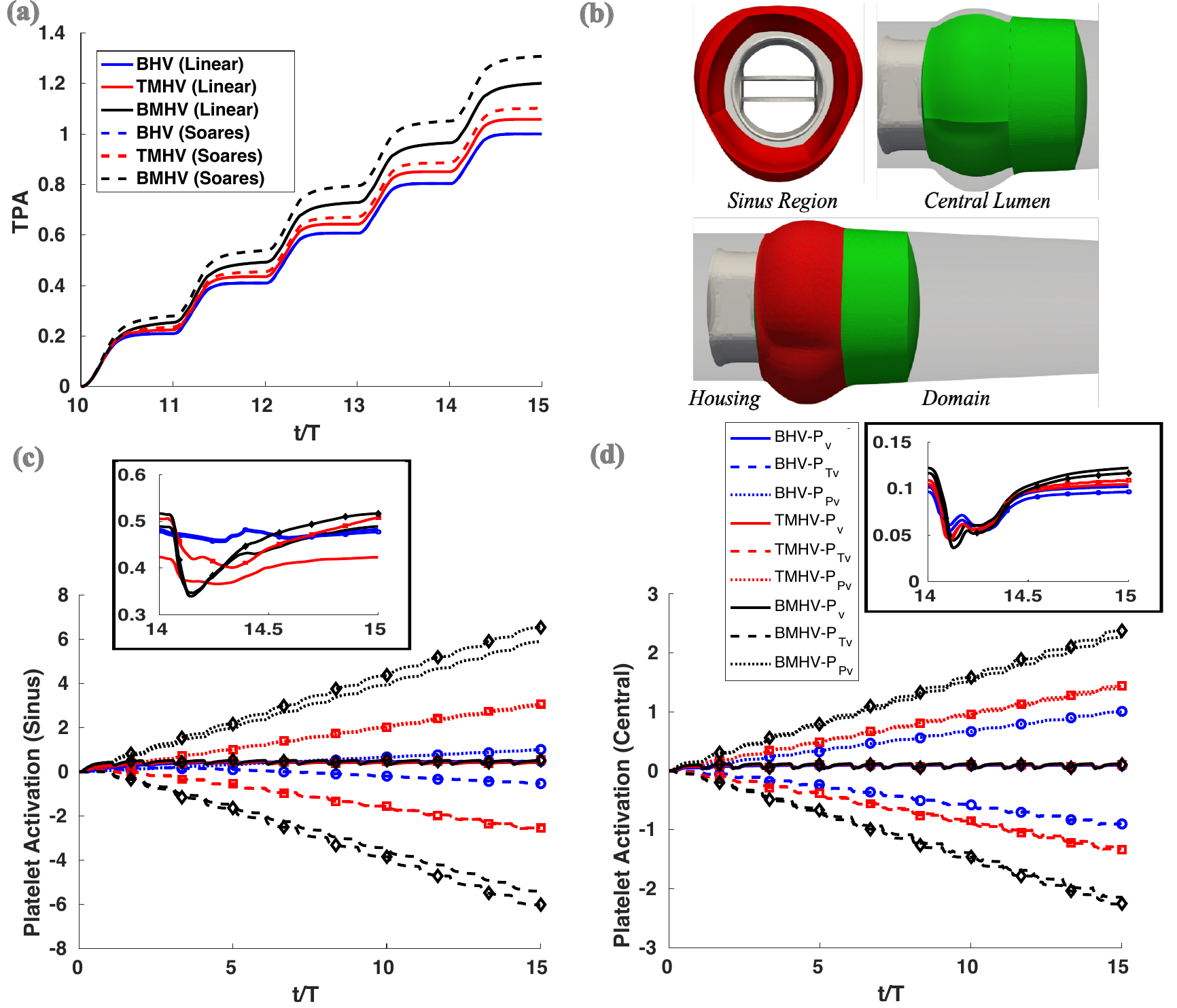}
			\caption{(a) Comparison of the TPA with results at the end of cycle 10 as datum (b) illustration of the sinus (colored red) and central lumen (colored green). Comparison of the volume integral of $P$, $P_T$ and $P_P$ is plotted in (c) for the sinus (inset for $P_V$ during cycle 15) and (d) for the central lumen (inset for $P_V$ during cycle 15). The solid lines represent the Linear model, whereas those with markers represent the Soares model.}
			\label{TPA}
		\end{center}
	\end{minipage}
\end{figure}

%%%%%%%%%%%%%%%%%%%%%%%%%%%%%%%

To answer this question, we first compare the TPA of the three valves, which is a measure of the time-integrated activation source of platelets, summed for the entire domain and normalized by that of the BHV at the end of diastole, in Figure \ref{TPA}a. At the end of the $15^{th}$ cardiac cycle, with the Linear model, the TPA is 20.07\% and 5.78\% higher in the BMHV and TMHV, respectively, compared to the TPA in BHV. With the Soares model, the TPA is 30.74\% and 10.21\% higher in the BMHV and TMHV, respectively, compared to the TPA in BHV, respectively. Regardless of the model, the BHV consistently shows lowest TPA (Figure \ref{TPA}a), which suggests that the BHV does not induce greater local shear-mediated platelet activation than the MHVs. Instead, it could be hypothesized that BHV promotes a greater retention of activated platelets in the aortic root, specifically in the aortic sinuses and near the arterial walls, limiting their convective transport.

To assess this hypothesis, we investigate the role of transport against local shear-induced activation in the sinus and central lumen regions, illustrated in Figure \ref{TPA}b, by plotting the temporal evolution of platelet activation levels tracked in a volume (i.e. $P_V$), those accumulated/washed away due to transport (i.e. $P_{T_V}$) and those that are locally activated due to shear stresses in the volume (\i.e. $P_{P_V}$), as defined in Equation (\ref{tracking-eq}), in Figures \ref{TPA}c and \ref{TPA}d, respectively. The results have been normalized with the $P_{P_V}$ of the BHV for each specific volume. Regardless of the model, the local shear-induced activation ($P_{P_V}$) and transport ($P_{T_V}$) of activated platelets reach a cyclic state in the sinuses (Figure \ref{TPA}c) and the central lumen (Figure \ref{TPA}d), resulting in a periodically stable volume integral of activated platelets $P_{V}$ in that region, for all valves. The $P_{P_V}$ in the MHVs remains significantly higher than that in the BHV throughout the cardiac cycle across the regions. At the end of the diastole, when a cyclic state is reached, the $P_{P_V}$ in the sinuses (Figure \ref{TPA}c) of the BMHV and the TMHV is respectively 490\% and 197.57\% higher than that in the BHV with the linear model. With the Soares model, the $P_{P_V}$ in the sinuses of the BMHV and the TMHV is respectively 554\% and 205\% higher than that in the BHV. However, following the early systolic acceleration through the end of systole, BHV demonstrates slightly higher $P_V$ than MHVs in the sinuses (refer to the inset in Figure \ref{TPA}c), despite the higher $P_{P_V}$ in MHVs. This observation is ascribed to the relatively lowest transport $P_{T_V}$ in BHV during systole. During diastole, $P_V$ in MHVs increases at a faster rate than BHV, primarily because their washout rate decreases while local shear-induced activation increases during early regurgitation. Owing to the near-compensatory evolution of $P_{P_V}$ and $P_{T_V}$ (Figure \ref{TPA}c) in MHVs, the resulting $P_V$ in the sinuses of the BMHV and TMHV differs only slightly from that of the BHV at the end of diastole. Specifically, with the Linear model, $P_{V}$ is 1.24\% higher for BMHV and 12.35\% lower for TMHV compared with BHV. With the Soares model, $P_{V}$ in the sinuses is 8.25\% and 6.5\% higher for BMHV and TMHV, respectively, than BHV.

Similar trends are observed in the central lumen (Figure \ref{TPA}d) at the end of the diastole, when a cyclic state is reached, the $P_{P_V}$ of the BMHV and the TMHV is respectively 126.65\% and 39.63\% higher than that in the BHV with the Linear model. With the Soares model, the $P_{P_V}$ in the central lumen of BMHV and the TMHV is respectively 137.75\% and 43.80\% higher than that in the BHV. The counterbalancing influence of $P_{T_V}$ on $P_{P_V}$ is more pronounced in the central lumen for all valves than was observed in the sinuses (Figure \ref{TPA}d). $P_V$ in the central lumen of the BMHV and the TMHV is 19.60\% and 2.54\% higher, respectively, than that in the BHV with the Linear model. With the Soares model, the $P_V$ in the central lumen of the BMHV and the TMHV is respectively 20.86\% and 13.05\%  higher than that in the BHV. 

A subtle observation is that $P_{T_V}$ remains consistently negative for all valves, once $P_V$ has stabilized periodically after the first few cardiac cycles, in the sinuses (Figure \ref{TPA}c) and the central lumen (Figure \ref{TPA}d), indicating a positive divergence of convective fluxes, that is, a net washout (a net flux out) of platelets with non-zero activation levels. This washout is most pronounced during the acceleration phase of the cardiac cycle across all valves, with BMHV exhibiting the highest washout rate, followed by TMHV and lowest in BHV. At the end of diastole after achieving a cyclic state, with the Linear model, the washout is quantitatively observed to be 10.45 and 2.40 times higher in BMHV while 4.94 and 1.45 times higher in TMHV compared to BHV in the aortic sinuses and central lumen, respectively. With the Soares model, the washout is observed to be 10.98 and 2.50 times higher in BMHV while 4.98 and 1.47 times higher in TMHV compared to BHV in the aortic sinuses and central lumen, respectively. These findings re-establish that BHV facilitates a weak transport of activated platelets from the aortic root, which explains the existence of stronger pockets of platelet activation levels in BHV than MHVs during systole (Figures \ref{linear-cont}a–d and \ref{soares-cont}a–d).

\begin{figure}[hbtp]
	\begin{minipage}{\textwidth}
		\begin{center}
			\includegraphics[width=1\textwidth]{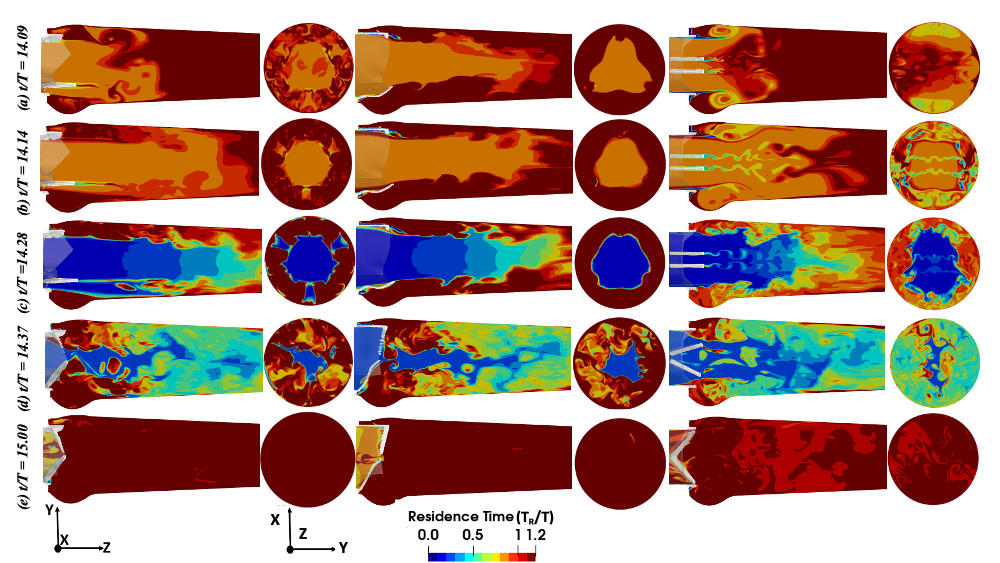}
			\caption{Eulerian residence time ($T_R$) in the TMHV (left), BHV (middle) and BMHV (right) at (a) early systole (b) peak systole (c) late systole (d) early regurgitation and (e) end of diastole. See Supplementary Movie 4.}
			\label{residencetime}
		\end{center}
	\end{minipage}
\end{figure}

Insights into the Eulerian $T_R$ field could further aid in understanding the observed trends in transport dynamics across the valves under consideration. We first evaluate and compare the planar contours of normalized residence time ($T_R/T$) in Figure \ref{residencetime}. During early systole (Figure \ref{residencetime}a), the incoming flow begins a washout (i.e. reduced $T_R/T$) of the aortic sinuses and central lumen in the MHVs while only the central lumen in the BHV. By peak systole (Figure \ref{residencetime}b), the orificial jets of all valves enable an efficient washout of the central lumen. The BMHV has wider side jets than the TMHV whereas no lateral jet exists in the BHV. Consequently, longest $T_R/T$ is observed in the sinus and near-arterial wall region beyond the sinus ridge of the BHV, followed by TMHV and BMHV (Figure \ref{residencetime}b-c). The TMHV and BHV close by early regurgitation (Figure \ref{residencetime}d), thereby promoting flow stagnation in the sinus and downstream aorta whereas the BMHV leaflets stay open until mid-diastole, permitting retrograde filling of the left ventricle. As a result, the aorta with BMHV shows shortest $T_R/T$ among all valves by early regurgitation (Figure \ref{residencetime}d) through the end of cycle (Figure \ref{residencetime}e). Notably, the $T_R/T$ at the end of diastole in the aorta with all valves is longest compared to the preceding cardiac instances due to a nearly stagnant flow, resulting from a prescribed near-zero regurgitation in our FSI simulations \citep{abbas2025closure}. 

%%%%%%%%%%%%%%%%%%%%%%%%%%%%%%%%

%%%%%%%%%%%%%%%%%%%%%%%%%%%%%%%

\begin{figure}[hbtp]
	\begin{minipage}{\textwidth}
		\begin{center}
			\includegraphics[width=0.6\textwidth]{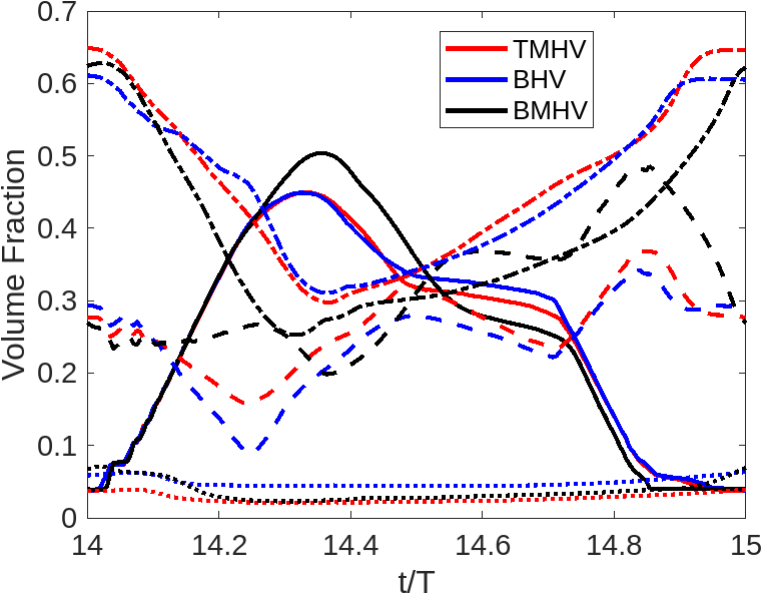}
			\caption{Comparison of volume fractions corresponding to four distinct ranges of $T_R/T$ across the valves.}
			\label{residencetimevf}
		\end{center}
	\end{minipage}
\end{figure}

A volumetric analysis of $T_R/T$ is illustrated in Figure \ref{residencetimevf}, which categorizes the total fluid volume of each valve into four fractions based on $T_R/T$, being $0 < T_R/T \leq 0.5$, $0.5 < T_R/T \leq 1$, $1 < T_R/T \leq 2$ and $T_R/T > 2$. At the start of the cycle ($t/T = 14$), more than 60\% of the fluid volume in each valve exhibits a $T_R/T$ in the range $1 < T_R/T \leq 2$, while more than 27\% lies in the range $0.5 < T_R/T \leq 1$, indicating that a substantial portion of the domain is occupied by fluid that has persisted from the previous cycle. BMHV demonstrates the highest volume fraction of $\approx 6.7\%$ in the range $T_R/T > 2$ compared to that of $\approx 5.9\%$ in BHV while $\approx 3.8\%$ in TMHV. For the sake of ease in description, the advancement of the cardiac time is classified into three phases: Phase 1 - early acceleration to end of systole ($ 14 < t/T < 14.34$), Phase 2 - early regurgitation to mid-diastole ($ 14.34 < t/T < 14.60$) and Phase 3 - mid-diastole to just before the end of diastole/cycle ($ 14.60 < t/T < 15$). As the Phase 1 begins, the volume fraction in the range $0 < T_R/T \leq 0.5$ increases at nearly the same rate across all valves, reflecting the forward entry of fresh fluid. Concurrently, the volume fractions decrease in the range $1 < T_R/T \leq 2$  and $T_R/T > 2$, while increase in the range  $0.5 < T_R/T \leq 1$ for all valves,  albeit most rapidly for the BMHV, suggesting a more efficient clearance of resident fluid. The BHV demonstrates a higher volume fraction in the ranges $1 < T_R/T \leq 2$ and $T_R/T > 2$, respectively being  $\approx 46\%$ and $\approx 4.5\%$ of the domain, among all valves for the later half of Phase 1 ($ 14.20 < t/T < 14.34$), showcasing longer residence of fluid due to poor washout. The BMHV exhibits a sustained increase in the range $0 < T_R/T \leq 0.5$ after the end of Phase 1 and into the early Phase 2 compared to TMHV and BHV, indicating a more prolonged washout phase for around 50\% of the domain, primarily because its leaflets have not fully closed after early regurgitation ($t/T > 14.37$), in contrast to TMHV and BHV. The three valves fully close during Phase 2 \citep{abbas2025closure}, latest by $t/T \approx 14.55$, after which the volume fraction corresponding to range $0 < T_R/T \leq 0.5$ continuously decreases, while that corresponding to $0.5 < T_R/T \leq 1$ and $1 < T_R/T \leq 2$ continuously increases for all of them. During Phase 3, the BMHV exhibits the smallest volume fraction in the range $1 < T_R/T \leq 2$, while highest in the range $0.5 < T_R/T \leq 1$ compared to TMHV and BHV. The BHV continues to record more than 6\% of the fluid domain in the range $T_R/T > 2$, being higher than MHVs during Phase 3. After the Phase 3 ends, the results match with those at the start of the cardiac cycle. Overall, the histograms of $T_R/T$ in Figure \ref{residencetimevf} complement the observations drawn from the planar contours shown in Figure \ref{residencetime}, that is, the MHVs provide a better washout than BHV.

%%%%%%%%%%%%%%%%%%%%%%%%%%%%%%%%

%%%%%%%%%%%%%%%%%%%%%%%%%%%%%%%

\begin{figure}[hbtp]
	\begin{minipage}{\textwidth}
		\begin{center}
			\includegraphics[width=1\textwidth]{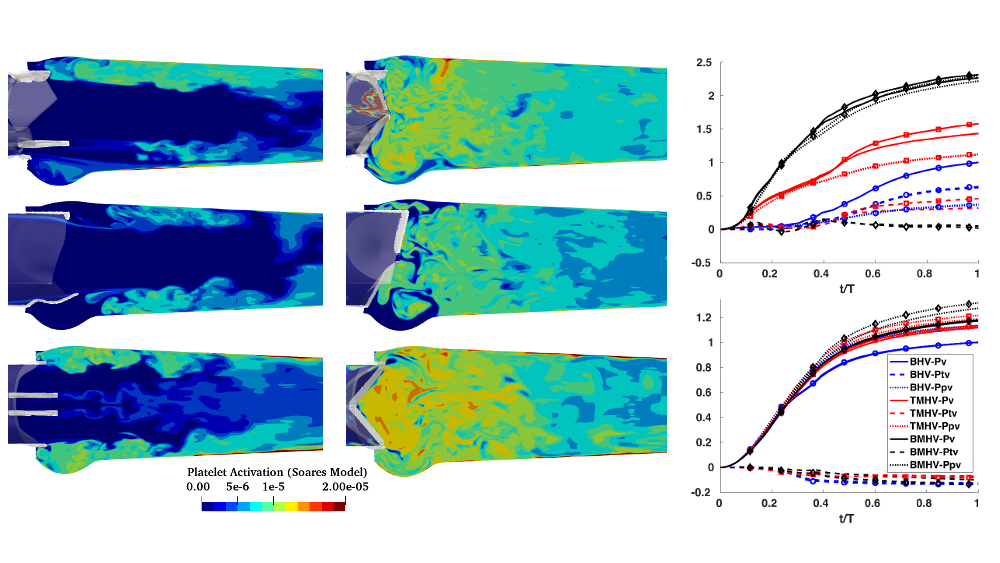}
			\caption{Platelet activation levels from the $1^{st}$ cardiac cycle for TMHV (top), BHV (middle) and BMHV (bottom) at (a) 100 ms after peak systole (b) mid diastole. A comparison of $P_V$, $P_{T_V}$ and $P_{P_V}$ from the first cardiac cycle is plotted in (c) for the sinus and (d) for the central lumen. The solid lines represent the results from the Linear model, whereas those with markers represent the results from the Soares model in (c) and (d).}
			\label{first-cycle}
		\end{center}
	\end{minipage}
\end{figure}

%%%%%%%%%%%%%%%%%%%%%%%%%%%%%%%%%%%%

It is worth noting that our analysis of the $1^{st}$ cardiac cycle reveals marked differences compared to the platelet activation results from the $15^{th}$ cycle. Around 100 ms after the peak systole of the $1^{st}$ cycle, BHV qualitatively shows weaker platelet activation levels than MHVs in the sinuses and near the arterial walls beyond the sinus ridge (Figure \ref{first-cycle}a). However, BHV sinuses begin to accumulate platelets with high activation levels after the onset of regurgitation (Figure \ref{first-cycle}b). Figures \ref{first-cycle}c quantitatively confirms these observations as during systole, the MHVs demonstrate a consistently positive $P_{T_V}$, indicating net accumulation in sinuses, with $P_{P_V}$ being significantly higher than $P_{T_V}$. In contrast, the BHV shows the lowest $P_{P_V}$ and a near-zero $P_{T_V}$ in the sinus (Figure \ref{first-cycle}c). After the onset of regurgitation, $P_{P_V}$ continues to dominate $P_{T_V}$ in MHVs, whereas $P_{T_{V}}$ for BHV exceeds $P_{P_V}$ by several folds in the sinuses (Figure \ref{first-cycle}c). The central lumen demonstrates a net washout for all valves (Figure \ref{first-cycle}d), with $P_{P_V}$ being significantly higher than $P_{T_V}$ during the $1^{st}$ cardiac cycle. 

Our results also assert that shear stress distributions and residence time contours do not directly correspond to platelet activation patterns in artificial heart valves, challenging the classical assumption of a direct correspondence \citep{borazjani2010high,haya2016effects,klusak2015high,hanafizadeh2016non,tango2025analysis}. For example, in the BHV, the sinus and the region near arterial wall exhibit the lowest shear stresses yet show the highest platelet activation among all valves. The same regions of the BHV although demonstrate longest $T_R/T$ compared to MHVs due to absence of lateral jets, however, by no means indicate a net washout of activated platelets as quantitatively shown by a negative $P_{T_V}$. Similarly, the immediate wake of the BMHV (covered by a vortex street) and its downstream region in the close proximity of the arterial walls demonstrate comparable magnitudes of viscous shear stresses, however, the platelet activation levels are higher in the latter region once a cyclic state is reached. It is, therefore, essential to carefully analyze the results with attention to both, local shear-induced platelet activation and the transport of activated platelets, rather than indiscriminately associating high shear stress regions with elevated platelet activation potential.

\section{Discussion}

In recent years, BHVs have become the preferred choice for aortic valve replacement over MHVs \citep{nitti2022numerical,puri2017bioprosthetic}. This shift is primarily driven by the high risk of clinical thrombosis in MHVs and the consequent reliance of the valve's recipient on lifelong oral anticoagulation, which increases the risk of hemorrhage. Nonetheless, BHVs also carry a risk of rare clinical or subclinical thrombosis \cite{rashid2023computed,sirois2011computational}, with more frequent occurrence in transcatheter BHVs than surgical ones \citep{tango2025analysis,makkar2022missing}. Our results might shed light on these differences, however, it is emphasized that platelets do not reach the activation thresholds set by the Linear model ($\int \tau dt$ being 35 $dynes/cm^{2} s$ \cite{hellums19941993}) or the Soares models (the platelet activation state, $\Delta$PAS being 1 \citep{soares2013novel}) at the end of the $15^{th}$ cycle in our simulations. The reference to terms as `activated platelets' or `platelets with high activation levels' denote platelets with non-zero activation levels, similar to previous studies \cite{bluestein2002free,min2014blood,hedayat2017platelet}.

Shear-induced platelet activation is widely recognized as a marker of thrombosis in BMHVs and has been shown to correlate strongly with the valve's hemodynamics \citep{bluestein2000vortex,bluestein2002free,dumont2007comparison,abbas2020num,hedayat2017platelet,hedayat2019comparison,yun2014computational}. Indeed, we see several fold more activation of platelets in sinus and central lumen regions of the MHVs compared to BHVs (Fig.~\ref{TPA}c,d). The high activation levels are balanced by high transport, which are again several folds higher than BHV (Fig.~\ref{TPA}c,d). This indicates that the larger regions with high shear stress, i.e., more platelets are exposed to shear stress, rather than residence time might be the main reason for thrombosis in MHVs.  
In contrast, our results infer that the subclinical thrombosis in BHV is mainly driven by prolonged residence times, owing to the absence of lateral jets that otherwise enhance washout in MHVs. This reduced clearance in the sinuses facilitates deposition of activated platelets in BHV, similar in effect to the stasis observed in the neosinuses of transcatheter BHVs, which has been shown to positively correlate with thrombus volume \citep{trusty2022role}. 

A review of previous studies reveals that most numerical simulations of platelet activation or blood damage potential in aortic valves have been limited to specific phases of the cardiac cycle \citep{alemu2007flow, hedayat2017platelet, bornemann2024relation} or, at most, a single cycle \citep{min2014blood, dumont2007comparison, hedayat2019comparison}. In contrast, our comparison of the $1^{st}$ and the $15^{th}$ cardiac cycles reveals striking differences in $P_V$, primarily driven by changes in $P_{T_V}$. While all valves demonstrate net accumulation of highly activated platelets (i.e., a positive $P_{T_V}$) in the sinuses at the end of the $1^{st}$ cycle, they later transition to a net washout state as the results become cycle-independent. The stark differences in the findings between the $1^{st}$ and $15^{th}$ cycles highlight the importance of simulating multiple cardiac cycles until a cyclic state is achieved, ensuring that the results are not influenced by cycle-to-cycle variability. In our study, a cyclic convergence was only observed after simulating 5 cycles for MHVs and 12 cycles for BHV, beyond which the time-averaged $P_V$ and volume-integrated $T_R$ varied by less than 1\% between the successive cycles. This feature may not be realistically attainable within the limitations of conventional Lagrangian frameworks to quantify platelet activation, since it will involve employing exorbitant computational resources.

Hedayat et al. \cite{hedayat2017platelet} demonstrated that the Linear model may underestimate platelet activation, as it neglects dynamic shear stress waveforms and associated nonlinear effects, which are accounted for in the Soares model. For this reason, the Soares model may offer a more accurate representation of platelet activation under artificial heart valve flow conditions. Our results show that platelet activation in MHVs is higher when predicted by the Soares model compared to the Linear model, using the BHV as a control, aligning well with the findings of Hedayat et al. \cite{hedayat2017platelet}. Nonetheless, both models yield the same relative trends, reinforcing that the primary conclusions of this study are robust and not dependent on the specific platelet activation model employed.

The limitations of this work include not resolving the hinge and leakage flows as was done in a previous study from our research group \citep{hedayat2019comparison} using overset grids to resolve such small areas.
Comparing the contributions from the bulk flow and hinge/leakage flows to the activation of platelets showed that despite higher shear stresses in the hinge/leakage area, the total activation induced by the bulk flow is several folds higher than that by the hinge/leakage flows because of the higher flow rate during systole exposes more platelets to elevated shear stress waveforms \cite{hedayat2019comparison}. Since the gaps and hinge recess were not resolved, a small opening/gap between the leaflets was incorporated to keep the flow connectivity between the left ventricle and aortic sides as required for FSI simulations \citep{abbas2025closure}.
This might overestimate regurgitation and washout of platelets with higher activation levels towards the left ventricle and underestimate the shear stresses during diastole. Furthermore, the coronary flows are neglected in this work, which may change the flow dynamics, especially for the BHV, in the sinuses.  Finally, Equation \ref{governing-eq} implicitly assumes that the number of resting platelets is significantly higher compared to activated platelets which is an acceptable assumption considering the large number density of platelets in blood, typically around $2.5 \times 10^5$ platelets/$mm^3$ \cite{montgomery2023clotfoam}.

\section{Conclusion}

This study provides a comprehensive comparative analysis of platelet activation levels in a TMHV, a BMHV and a BHV. A novel aspect of the work is its focus on identifying whether platelets with high activation levels are activated locally within the sinuses and central lumen, or are accumulated from elsewhere via flow-driven transport. By simulating 15 cardiac cycles and reaching the cyclic state, using both Linear and Soares models, we demonstrate that the volume integral of the time-integrated platelet activation in any region ($P_V$) stabilizes once the transport fluxes ($P_{T_V}$) and activation source ($P_{P_V}$) in that region reach a periodic balance. 
 
The BHV induces the lowest local shear-mediated activation among all valves. With the Linear model, the TPA (a measure of total platelets that were activated) was observed to be 20.07\% higher in BMHV while 5.78\% higher in TMHV with respect to BHV at the end of the $15^{th}$ cardiac cycle. With the Soares model, the TPA was observed to be 30.74\% higher in BMHV while 10.21\% higher in TMHV with respect to BHV. The TPA in the current commercially preferred mechanical prosthesis, the BMHVs, is therefore around 14.30\% higher than the TMHV with the Linear model, and around 20.50\% higher with the Soares model. Interestingly, BHV demonstrates the weakest washout of the sinuses and vicinity of the arterial wall, evidenced by the prolonged residence time in these regions compared to MHVs. BMHV demonstrates the highest washout of the sinus and central lumen among all valves, followed by TMHV and lowest in BHV. Consequently, the amount of activated platelets remaining in any region, quantified by $P_V$, in MHVs is similar to BHV. Notwithstanding the choice of activation model, the findings remain consistent.

\appendix

\section{Governing Equations in Curvilinear Coordinates}\label{app:num-method}

The previous flow simulations \cite{abbas2025closure} were performed on a curvilinear grid using the CurvIB method \citep{borazjani2008curvilinear,asadi2023contact}. Therefore, the transport equations (Equation ~\ref{governing-eq} and Equation \ref{residencetime-eq}) need to be solved on a curvilinear grid. Contextually, these transport equations for a scalar Q could be written as:

\begin{equation}
\displaystyle\frac{\partial Q}{\partial t} + C(Q) = \frac{1}{Pe}D(Q) + S(\tau)
  \label{curvilinear-eq}
\end{equation}
where, $C(Q)$ and $D(Q)$ are the advection and diffusion operators, respectively. After the curvilinear coordinate transformation ($\zeta^r=\zeta^r(x_1, x_2,x_3)$ where $r=1,2,3$), the advection operator could be written as follows, in index notation:

\begin{equation}
C(Q) = \nabla \cdot(uQ)= J \left [ \frac{\partial}{\partial \zeta^r}\left( \frac{V^rQ}{J} \right) \right]
  \label{adv-op}
\end{equation}

where, $J$ is the Jacobian of the geometric transformation such that $J=\displaystyle\frac{\partial (\zeta^1,\zeta^2,\zeta^3)}{\partial (x^1,y^2,z^3)}$, $V^r$ represent the contravariant velocity components defined as $V^r=u_j \zeta^r_{x_j}$, and $\zeta^r_{x_j}=\partial \zeta^r/\partial x_j$. The diffusion operator takes the following form:

\begin{equation}
D(Q) = \nabla^2 Q = J \left [ \frac{\partial}{\partial \zeta^r}\left( \frac{g^{rm}}{J} \frac{\partial Q}{\partial \zeta^m} \right)\right]
  \label{diff-op}
\end{equation}

\noindent{where, $g^{rm}$ is the contravariant metric tensor, such that $g^{rm}=\zeta_{xq}^r\zeta_{xq}^m$}. The source terms that do not contain spatial derivatives and remain the same in curvilinear coordinates as cartesian ones. 

Note that we use the conservative form of equations, and by our finite volume and staggered arrangement of variables ($V^r$ on cell faces while $Q$ at cell centers), we satisfy the discretized conservation equations to machine zero, i.e., fully satisfy the conservation of species to machine zero.

%\clearpage
\bibliographystyle{elsarticle-num}
\bibliography{Ref}

\end{document}